\documentclass[11pt, letterpaper]{article}

\usepackage{graphicx}
\usepackage{amsmath}
\usepackage{amssymb}
\usepackage{setspace}
\usepackage{epstopdf}
\usepackage{color}
\usepackage[letterpaper,left=1in,right=1in,top=1in,bottom=1in]{geometry}
\usepackage[]{algorithm2e}
\usepackage{multirow}
\usepackage{longtable}
\usepackage{rotating}
\usepackage{mathrsfs}

\newtheorem{rmk}{Remark}

\newtheorem{alg}{Algorithm}

\newcommand{\m}{\mathbb}

\title{\Large \bf A consensus-based approach for parameter and state estimation of agro-hydrological systems }

\author{
\centerline{\normalsize Xunyuan Yin$^{a}$, Song Bo$^{a}$, Jinfeng Liu$^{a}$, Biao Huang$^{a}$
}
\vspace{5mm}\\
\centerline{\small $^{a}$Department of Chemical \& Materials Engineering, University of Alberta,}\\
\centerline{\small Edmonton, AB T6G 1H9, Canada}}
\allowdisplaybreaks

\begin{document}

\date{}
\maketitle
\setstretch{1.5}

\begin{abstract}
The development of advanced closed-loop irrigation systems requires accurate soil moisture information. In this work, we address the problem of soil moisture estimation for the agro-hydrological systems in a robust and reliable manner. A nonlinear state-space model is established based on the discretization of the Richards equation to describe the dynamics of agro-hydrological systems. We consider that model parameters are unknown and need to be estimated together with the states simultaneously. We propose a consensus-based estimation mechanism, which comprises two main parts: 1) a distributed extended Kalman filtering algorithm used to estimate several model parameters; and 2) a distributed moving horizon estimation algorithm used to estimate the state variables and one remaining model parameter. Extensive simulations are conducted, and comparisons with existing methods are made to demonstrate the effectiveness and superiority of the proposed approach. In particular, the proposed approach can provide accurate soil moisture estimate even when poor initial guesses of the parameters and the states are used, which can be challenging to be handled using existing algorithms.
\end{abstract}


\section{Introduction}

Water scarcity has been one of the greatest global issues. According to a report of the United Nations \cite{UN_report}, approximately 70\% of fresh water is consumed by agriculture-related activities. Currently, open-loop irrigation systems are still widely used, in which the irrigation amount is determined based on experience rather than the actual soil moisture information of a field. The current
water-use efficiency in irrigation is only about 50\% -- 60\% \cite{Fischer2007}. To achieve water sustainability, irrigation efficiency needs to be substantially improved. One promising approach is to develop closed-loop irrigation systems by incorporating real-time feedback information (e.g., soil moisture) in an irrigation decision making process \cite{MAO2018Irrigation}.
The development of advanced irrigation systems requires an accurate agro-hydrological model and real-time soil moisture information at different points of the soil profile, both of which can be challenging to obtain directly \cite{Nahar2018CCE}. Parameter and state estimation serves as a natural solution to this type of problems. In this work, the objective is to propose a systematic parameter and state estimation approach for agro-hydrological systems based on a dynamic model constructed using the Richards equation \cite{richards1931capillary}.

The Richards equation has been commonly used to describe soil water dynamics, and some parameters involved in the Richards equation that are related to soil properties need to be estimated accurately. One solution to this parameter estimation problem is to conduct soil water retention curve fitting in a lab environment using measurements collected in an offline manner \cite{van1980closed}. However, these parameters may be time-varying, so that this offline-based solution may not be able to provide sufficiently accurate parameter estimates as needed for online soil moisture estimation. Also, it can be expensive or at least time-consuming to collect samples and send them to a lab for offline analysis. In another line of work, optimization-based solutions were formulated on the basis of minimizing suitable objective functions that express the difference between sensor measurements and the predicted values \cite{Hwang2003,Russo1991}. This type of methods can only be used to estimate the model parameters, yet not the soil moisture. Moreover, these methods in general require fairly accurate initial guesses of the parameters, and were not implemented for online evaluation.

Problems related to soil moisture estimation have also been investigated \cite{Nahar2018CCE,Reichle2002,Erdal2015,Zhang2017,lu2011}. For example, in \cite{Walker2001b}, a Kalman filter was developed to estimate water content in the soil based on the linearization of a simplified soil profile model.
In \cite{Sahoo2019aiche}, a systematic approach was proposed to address optimal sensor placement for moisture estimation. In our recent work \cite{Bo2020}, a simultaneous parameter and state estimation method was developed for agro-hydrological systems by incorporating the unknown parameters as states. Using this approach, the capillary pressure head and a few model parameters can be estimated. However, soil moisture is yet to be reconstructed, and the estimation accuracy is highly dependent on the quality of the initial guesses. In particular, when a less accurate initial guess is supplied, the estimates may not converge to the actual values or even diverge.

Based on the above observations, we aim to propose a more powerful and robust solution to the problem of simultaneously estimating the unknown model parameters and soil moisture. In particular, the new solution is expected to be able to provide online accurate estimates of all the unknown parameters and soil moisture, and can handle poor initial guesses of the parameters and/or the soil moisture. The consensus-based distributed architecture can be a promising candidate to address the above limitations.
A consensus-based estimation scheme incorporates local estimators that can collaborate with each other to provide overall more reliable estimates. Specifically, each local estimator takes advantage of local sensor measurements and communicates with other estimators to provides updated estimates of the states/parameters, aiming to reach an agreement on the estimates with the other estimators gradually \cite{2008IEEETSP,Bandyopadhyay2014ACC}.
In \cite{OlfatiSaber2007CDC}, a consensus-based distributed Kalman filtering algorithm was proposed for state estimation of linear sensor networks. In \cite{Sahu2016}, a consensus-based distributed estimation scheme was proposed for estimating unknown parameters of nonlinear functions. A distributed moving horizon estimation scheme was developed for constrained systems using consensus \cite{farina2012distributed}. Moreover, the use of a distributed structure can lead to improved fault tolerance, organizational and maintenance flexibility of estimation schemes \cite{Daoutidis2016JPC,Jiang2019IECR,CSML12CCE,Daoutidis2019AIChE,Yin2019AIChE,
Mhaskar2018,Tang2019CJCE,Pourkargar2019iecr}.


In this work, we address the problem of model parameter and soil moisture estimation for agro-hydrological systems. In particular, we consider that the model parameters related to soil properties are unknown and good initial guesses are unavailable. A consensus-based distributed architecture is exploited, and simultaneous parameter and soil moisture estimation is addressed by proposing two distributed estimation algorithms based on consensus, which provide more reliable and robust estimates of the parameters and the states. Specifically, a nonlinear state-space model is established to describe the dynamics of the systems, and the output measurement equation is established based on the soil-water retention relationship. A distributed extended Kalman filtering (DEKF) scheme is proposed to estimate the parameters involved in the output measurement equation. A distributed moving horizon estimation (DMHE) scheme that can explicitly address constraints on states is proposed to estimate the soil moisture at different depths and a remaining model parameter. A consensus-based distributed mechanism that coordinates the communication between the DEKF and the DMHE, and governs the evaluation of the two schemes is also developed. Extensive simulations are carried out to demonstrate the efficacy of the proposed mechanism, and the superiority of the proposed method is illustrated in comparison to a centralized moving horizon estimation design. In particular, the results confirm that the proposed mechanism can substantially improve the accuracy of the estimates and the robustness of both state and parameter estimation against external noise and inaccurate initial guesses.
It can also help avoid the divergence of the estimates from the actual values, which may be encountered using centralized designs.
This point will be further discussed using the results in Section~\ref{section:results}.

\section{System description and problem formulation}

\subsection{Notation}
$\mathbb K_{+}$ represents a set that contains all non-negative integers. $\text{diag}\left(v\right)$ is a diagonal matrix of which the main diagonal consists of the elements of the vector $v$. ${I}_{n}$ represents an identity matrix of size $n$.
$\mathbb Z$ is the set of non-negative integers.
$\left\|v\right\|^2_Q$ denotes the square of the weighted Euclidean norm of vector $v$, which is computed as $\left\|v\right\|^2_Q=v^TQv$ where $Q$ is a positive definite weighting matrix.
$\left\{s\right\}^{d}_{c}$ represents a column vector containing a sequence of $s$ from time $k=c$ to $k=d$; that is, $\left\{s\right\}^{d}_{c}=\left[s_c,s_{c+1},\ldots,s_d\right]^{\text{T}}$.

\subsection{System model description}
We consider an agro-hydrological system that describes the interaction among the soil, crops and the atmospheric environment within the hydrological cycle. A schematic of the considered system is presented in Figure~\ref{fig:agro:schematic}. The input to this system is the water flow entering/exiting the soil, which may consist of rainfall, irrigation, water extraction by plant roots and evaporation, drainage and water runoff.
In this system, only the vadose zone of the soil is considered. In addition, the following assumptions are made: (1) soil properties are horizontally homogeneous; (2) irrigation is applied to the surface of the field uniformly. Consequently, we take into account the vertical hydrological dynamics in the system, and the dynamics of the water flows are characterized by the Richards equation \cite{richards1931capillary} in the following form:

\begin{figure}
    \centering
    \includegraphics[width=0.5\textwidth]{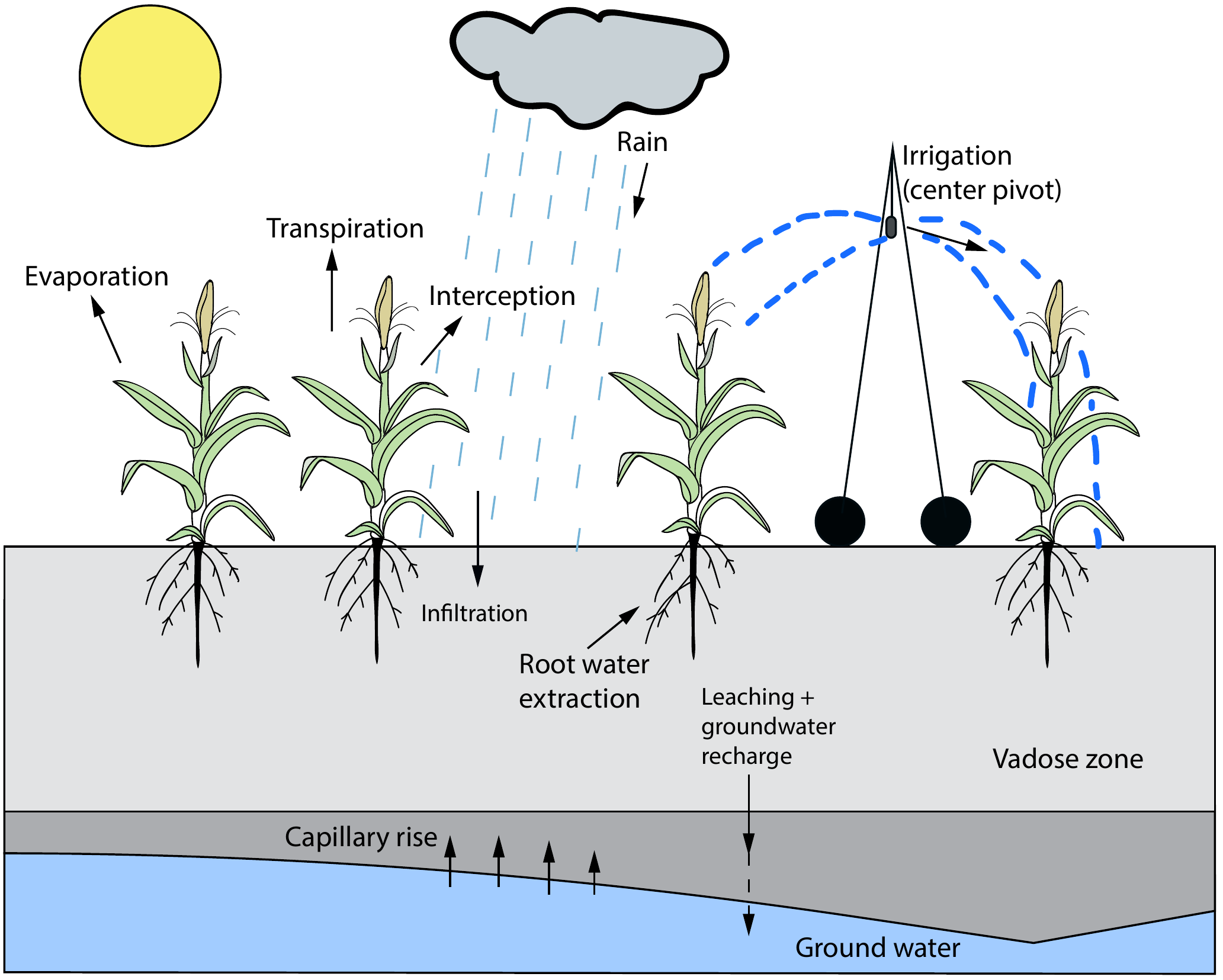}
    \caption{A schematic of the considered agro-hydrological system.}
    \label{fig:agro:schematic}
\end{figure}

\begin{equation}\label{eq:richards}
     \dfrac{\partial \theta}{\partial t} = C(h) \frac{\partial h}{\partial t} = \frac{\partial}{\partial z} \left[K(h)\left(\frac{\partial h}{\partial z}+1\right)\right]
\end{equation}
where $\theta$ (m/m) denotes the moisture content in the soil, $z$ (m) indicates the vertical position inside the soil, $h$ (m) is the capillary pressure head (also known as soil water pressure head), and $K$ (m/s) represents the hydraulic conductivity, which is dependent on $h$.
The dependence of the hydraulic conductivity ($K(h)$) on the pressure head ($h$) can be characterized using the following equation \cite{mualem1976new}:
\begin{equation}\label{eq:Kh}
K(h)=
 K_{sat}S_e^\lambda\left[1-\left(1-S_e^{\frac{1}{m}}\right)^m\right]^2
\end{equation}
where $m=1-1/n$, $\lambda$, $\alpha$ and $n$ are empirical parameters reflecting the characteristics of the soil, $K_{sat}$ denotes the saturated hydraulic conductivity, and $S_e$ denotes the relative saturation computed through the following equation:
\begin{equation*}
S_e = \frac{\theta(h) - \theta_r}{\theta_s-\theta_r}
\end{equation*}
where $\theta_s$ denotes the saturated soil water content, and $\theta_r$ denotes the residual soil water content.
The relation between moisture content $\theta(h)$ and the pressure head can be characterized by the soil-water retention equation \cite{van1980closed}:
\begin{equation}\label{eq:retention}
\theta(h) =
     \left(\theta_{s}-\theta_{r}\right)\left(1+(-\alpha h)^n\right)^{-\left(1-\frac{1}{n}\right)}+\theta_{r}
\end{equation}
Further, the capillary capacity (i.e., $C(h)$ in Eq. (\ref{eq:richards})) can be calculated as follows \cite{van1980closed}:
\begin{equation}\label{eq:capacity}
    C(h) = n \alpha \left(\theta_{s}-\theta_{r}\right)\Big(1-\Big(\frac{1}{n}\Big)\Big)\left(-\alpha h\right)^{n-1}\left(1+\left(-\alpha h\right)^n\right)^{-\left(2-\frac{1}{n}\right)}
\end{equation}


\subsection{Discretization of the Richards equation}

As shown in (\ref{eq:richards}), the Richards equation is a nonlinear partial differential equation (PDE) with respect to variables in both time and spatial coordinates. In this work, finite difference is exploited to provide a numerical approximation of the considered PDE, and both time discretization and space discretization of the model are conducted. Following the discretization in \cite{Bo2020}, two-point forward difference is employed to obtain an approximation of the derivative of the function with respect to time, which is carried out as follows:
\begin{equation}\label{eqn:fdm_time}
	\frac{\partial h_{i}\left(k\right)}{\partial t}\approx\frac{h_{i}\left(k+1\right)-h_{i}\left(k\right)}{\Delta t}
\end{equation}
where the subscript $i\in\left\{1,2,\ldots,n_x\right\}$ represents the index of each discretized node, with $n_x$ denoting the number of nodes.
$\Delta t$ represents the size of the time step used in discretization.

In addition, two-point central difference is employed to approximate the derivatives with respect to space as follows:
\begin{equation} \label{eqn:fdm_space}
\begin{aligned}
&\frac{\partial}{\partial z}\left(K_{i}\left(h\left(k\right)\right)\left(\frac{\partial h_{i}\left(k\right)}{\partial z}+1\right)\right) \\
&~~~~~\approx \frac{K_{i-\frac{1}{2}}\left(h\left(k\right)\right)\left(\frac{h_{i-1}\left(k\right)-h_{i}\left(k\right)}{\frac{1}{2}(\Delta z_{i-1}+\Delta z_{i})}+1\right)-K_{i+\frac{1}{2}}\left(h\left(k\right)\right)\left(\frac{h_{i}\left(k\right)-h_{i+1}\left(k\right)}{\frac{1}{2}(\Delta z_{i}+\Delta z_{i+1})}+1\right)}{\Delta z_{i}}
\end{aligned}
\end{equation}
where $\Delta z_i$ and $\Delta z_{i-1}$ represent the sizes of the space steps for two consecutive nodes used in discretization, and are made identical in this work.
In (\ref{eqn:fdm_space}), central difference is employed, and the subscripts $i-\frac{1}{2}$ and $i+\frac{1}{2}$ for the hydraulic conductivity $K$ represent the average of the two neighboring nodes. The hydraulic conductivity values at center points of two neighboring nodes are approximated by $K_{i - \frac{1}{2}}(h) \approx K(\frac{h_{i-1}+h_{i}}{2})$ and $K_{i + \frac{1}{2}}(h) \approx K(\frac{h_{i}+h_{i+1}}{2})$.

Based on the model in (\ref{eq:richards}) as well as the approximations in (\ref{eqn:fdm_time}) and (\ref{eqn:fdm_space}), a discrete-time finite difference equation for node $i$ is established as follows:
\begin{equation}\label{eqn:fdm_dt}
h_{i}(k+1) \approx h_{i}(k) + \Delta t \frac{K_{i-\frac{1}{2}}(h(k))\left(\frac{h_{i-1}(k)-h_{i}(k)}{\frac{1}{2}(\Delta z_{i-1}+\Delta z_{i})}+1\right)-K_{i+\frac{1}{2}}(h(k))\left(\frac{h_{i}(k)-h_{i+1}(k)}{\frac{1}{2}(\Delta z_{i}+\Delta z_{i+1})}+1\right)}{\Delta z_{i}c_{i}(h(k))}
\end{equation}
In this work, we apply the Neumann boundary condition to the top and bottom boundaries of the system, and it is obtained that:
\begin{equation} \label{eqn:tbc}
\begin{aligned}
	\left.\frac{\partial h\left(k\right)}{\partial z}\right|_{t} &= -1 - \frac{q_{t}(k)}{K\left(h\left(k\right)\right)}\\[0.2em]
	\left.\frac{\partial h\left(k\right)}{\partial z}\right|_{b} &= 0
\end{aligned}
\end{equation}
where the subscript $t$ indicates the top boundary while $b$ indicates the bottom boundary, and $q_{t}$ $(m/s)$ is the rate of overall water flow entering the soil, which is considered as the known input to the system,
and free drainage boundary condition is applied to the bottom layer.

{\color{red}
}

A discrete-time state-space model is established as in the following compact form:
\begin{equation}\label{eq:state:space}
x(k+1)= f\left(x(k), \bar\beta, u(k)\right) \\
\end{equation}
where $x=h\in\mathbb R^{n_x}$ is the vector of capillary pressure head at all the discretized nodes, $\bar\beta = \left[K_{sat}, {\beta}^{\text{T}}\right]^{\text{T}}$ is a vector of the model parameters with $\beta = \left[\theta_s, \theta_r, \alpha, n\right]^{\text{T}}$, $u = q_t \in\mathbb R^{n_1}$ represents the manipulated input to the soil, and $f(\cdot,\cdot,\cdot)$ is a vector field obtained based on the discretization of (\ref{eq:richards}) and the expressions of $K(h)$ in (\ref{eq:Kh}) and $C(h)$ in (\ref{eq:capacity}). From a practical point of view, we consider that all the model parameters in $\bar\beta$ are unknown.

\subsection{Problem formulation}
In the soil profile, there are $n_y$ ($n_y<n_x$) sensor nodes.
At each of these $n_y$ sensor nodes, both a moisture sensor and a tensiometer (i.e., one common type of soil water pressure head sensor) are deployed. This type of pressure sensors has been widely used in agricultural activities \cite{Walter1993}.
At every sampling time, each active moisture/soil water pressure head sensor provides a soil moisture/capillary pressure head measurement for the corresponding node.


Consequently, a compact state-space model for each sensor node is represented as follows:
\begin{subequations}\label{eq:state:space:local}
\begin{align}
x(k+1)&= f\left(x(k),\bar\beta, u(k)\right) + w(k)\label{eq:state:space:local:1}\\
y_j(k) &= g_j\left(x(k),\beta\right) + v_j(k)\label{eq:state:space:local:2}
\end{align}
\end{subequations}
where $y_j$ is the soil moisture measurement at the $j$th sensor node, $j\in\mathbb I=\left\{1,\ldots,n_y\right\}$,
$w\in\mathbb R^{n_x}$ denotes the additive system disturbances,
and $v_j$ is the measurement noise at the $j$th sensor node. (\ref{eq:state:space:local:2}) accounts for the output measurement function at the $j$th sensor node,  and $g_j$ is a nonlinear equation formed based on (\ref{eq:retention}) such that:
\begin{equation}\label{eq:retention:nodei}
g_j( x,\beta) :=
     \left(\theta_{s}-\theta_{r}\right)\left(1+\left(-\alpha x^j\right)^n\right)^{-\left(1-\frac{1}{n}\right)}+\theta_{r}
\end{equation}
where $x^j$ is the capillary pressure head at sensor node $j$, $j\in\mathbb I$. Note that superscript $j$ is used in $x^j$ so that it is distinguished from $x_i$ that represents the $i$th element of the pressure head vector $x$.

The ultimate goal is to obtain a good estimate of the soil moisture at each of the $n$ nodes by taking advantage of the available sensor measurements.
It is seen that the pressure formulation of the Richards equation in (\ref{eq:richards}) with $C(h) \frac{\partial h}{\partial t}$ on the left-hand-side is used, and the states in model (\ref{eq:state:space:local}) are the capillary pressure head. Therefore, our objective is first to estimate $\beta$ and provide an online estimate of $h(k)$ for $k=0,1,\ldots$, then an estimate of the soil moisture profile at the discretized nodes can be obtained based on the estimates of $\beta$ and $h(k)$, resorting to the soil-water retention equation in (\ref{eq:retention}).


\begin{figure}[t]
\centerline{\includegraphics[width=0.53\textwidth]{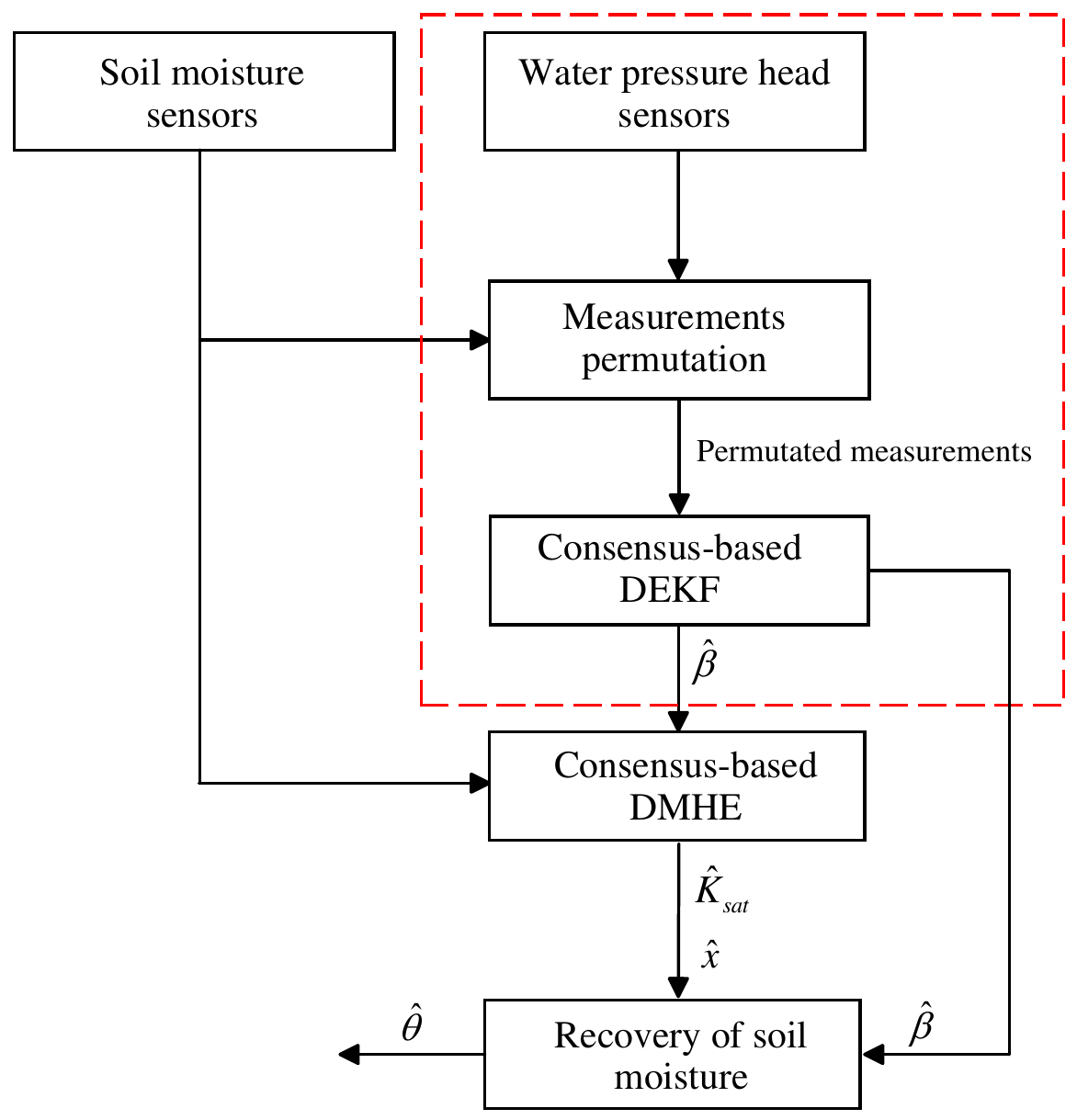}}
\caption{A flowchart of the consensus-based estimation mechanism (only the steps in the red dashed box are performed when the parameters are not available for state estimation).}
\label{flowchart_propose_design}
\end{figure}
\section{Architecture of the proposed mechanism}

Due to the unavailability of the five model parameters involved in $\bar\beta = \left[K_{sat}, {\beta}^{\text{T}}\right]^{\text{T}}$, it is necessary to also estimate these parameters as needed for the estimation of the soil moisture. Based on this consideration, we propose a consensus-based distributed estimation mechanism, which comprises a distributed extended Kalman filter (DEKF) scheme for parameter estimation for the model in (\ref{eq:retention}) and a distributed moving horizon estimation (DMHE) scheme for state and parameter estimation based on (\ref{eq:state:space:local}). The DEKF is used to estimate four unknown parameters involved in (\ref{eq:retention}), while the DMHE is exploited to estimate the remaining parameter and the system states. A flowchart that sketches the proposed mechanism is given in Figure~\ref{flowchart_propose_design}. At the initial stage when the estimates of the five parameters are not available, both the soil moisture sensors and the capillary pressure head sensors are utilized to collect measurements. The consensus-based DEKF scheme gives estimates of the parameters in $\beta$ (denoted by ${\hat{ \beta}}$) in a recursive manner based on Eq.~(\ref{eq:retention}) by taking advantage of both the moisture measurements and the pressure measurements.
After the estimates of the parameters in $\beta$ have converged, they will be sent to the consensus-based DMHE mechanism for estimating the remaining parameter $K_{sat}$ as well as all the states in terms of soil water pressure head every sampling time. The estimates of the capillary pressure head at all the discretized nodes and the estimate $\hat{ \beta}$ are substituted into the soil-water retention equation to obtain estimates of the soil moisture information at all the discretized nodes. The key steps involved in the entire consensus-based mechanism will be summarized in Section \ref{section:recovery}.

\begin{figure}[t]
\centerline{\includegraphics[width=0.72\textwidth]{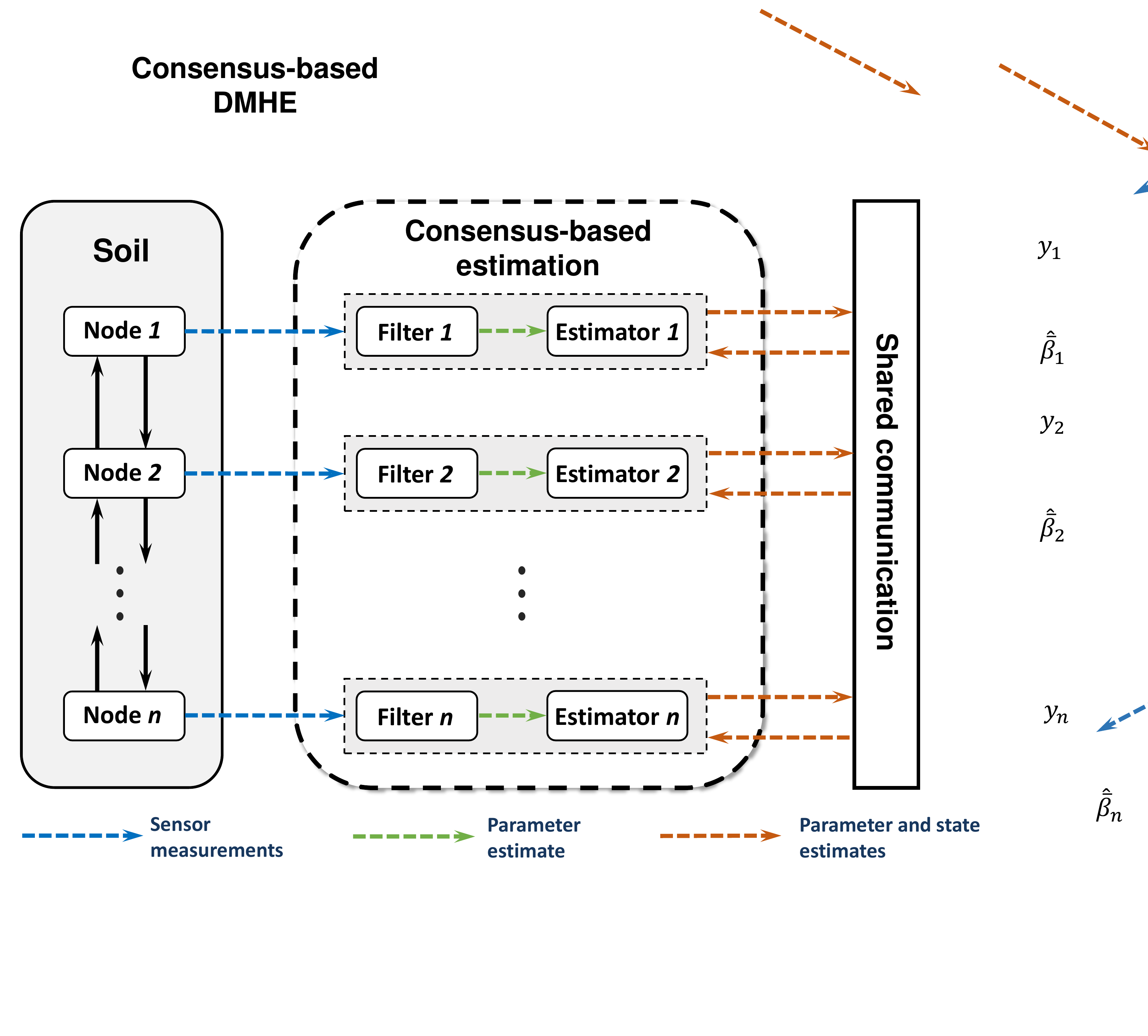}}
\caption{A schematic of the consensus-based distributed estimation mechanism.}
\label{schematic_distributed_design}
\end{figure}

A schematic of the consensus-based distributed state estimation mechanism is presented in Figure~\ref{schematic_distributed_design}. The entire distributed mechanism comprises two schemes: the consensus-based DMHE and the consensus-based DEKF. The DMHE contains $n_y$ local estimators and the DEKF contain $n_y$ local filters; that is, each sensor node has one local estimator and one local filter.
In Figure~\ref{schematic_distributed_design} and the remainder of the paper, each filter represents a local filter of the DEKF scheme, and each estimator represents a local estimator of the DMHE scheme.
The sensors at each node collect sensor measurements, which are sent to the estimator/the filter corresponding to the same sensor node. Each  estimator/filter communicate with the other estimators/filters to exchange the most recent estimates at each step. The parameter estimates provided by each filter will be used by the estimator for the same subsystem for state estimation.
The detailed designs of the two consensus-based estimation schemes and the corresponding local estimators will be introduced in detail in the following sections.

\section{Consensus-based DEKF for parameter estimation}

In this section, we propose a consensus-based distributed extended Kalman filtering (DEKF) design that comprises local filters coordinating with each other through communication. The DEKF is to be used for estimating the unknown parameters in the soil-water retention equation, i.e., $\theta_s$, $\theta_r$, $\alpha$ and $n$ in (\ref{eq:retention}).

 Within the DEKF scheme, one filter is developed for each sensor node in the soil. At each sampling time,  both the soil moisture and the capillary pressure measurements for each node (which are called a measurement pair in the remainder) are sent to the filter of the DEKF for the same node. Also, all the currently available sensor measurements are randomly permutated and are then used by the local DEKF filters. Each local filter of the DEKF is evaluated iteratively and eventually provides an estimate of $\hat{\beta}$ in the last iteration step during each sampling time. In what follows, we first present the design of the filters of the DEKF. Then, we introduce an implementation algorithm for the consensus-based DEKF to estimate the four unknown parameters involved in $\beta$.

\subsection{Local estimator design}
The $i$th filter, $i = 1\ldots,n_y$, of the consensus-based DEKF is designed as in the following form for $l=0,\ldots,k$:
\begin{equation}\label{eq:local:estimator}
{\hat{\beta}}^{(k)}_i(l+1)= {\hat{\beta}}^{(k)}_i(l) + L_i(l)\Big(y_i(l)-  g_i\big(x(l),{\hat {\beta}}^{(k)}_i(l)\big) \Big)- \mu(l) \Big({\hat {\beta}}^{(k)}_i(l)-\sum_{j\in\mathbb I\setminus\left\{i\right\}} \frac{1}{n_y-1} \hat {\beta}^{(k)}_{j}(l)\Big)
\end{equation}
where
\begin{subequations}\label{eq:local:estimator:2}
\begin{align}
L_i(l)&= P_i(l)H_i^{\text{T}}(l)\left(H_i(l)P_i(l)H_i^{\text{T}}(l)+R\right)^{\text{-1}}\label{eq:local:estimator:2:a}\\[0.5em]
H_i(l) &= \frac{\partial g_i\big(x,{\beta}\big)}{\partial\beta}\bigg|_{x(l),{\hat{ \beta}}_i^{(k)}(l)}\label{eq:local:estimator:2:b}\\[0.5em]
P_i(l+1) &= \left({\color{black}\lambda(l)}+1\right)\left(P_i(l)-L_i(l)H_i(l)P_i(l)+Q\right)\label{eq:local:estimator:c}
\end{align}
\end{subequations}
In (\ref{eq:local:estimator}), ${\hat{\beta}}^{(k)}_i(l+1)$ denotes the estimate of $\beta$ given by the $i$th filter in the $l$th step at sampling time $k$, $y_{i}(l)$ is the $l$th moisture measurement for node $i$ after random permutation of all the available measurements at sampling time $k$, $Q$ and $R$ are positive-definite matrices that are made identical for each step $k$, $k\in\mathbb Z$, and $\mu>0$ and $\lambda>0$ are two time-varying tuning parameters that need to be made positive.

In filter (\ref{eq:local:estimator}), the second term on the right-hand-side adapted from \cite{alessandri2007recursive} is a correction term based on the innovation corresponding to the $i$th sensor node. The calculation of the correction gain $L_i$ is similar to the standard EKF algorithm, except that in (\ref{eq:local:estimator:c}), an additional coefficient $\lambda$ is introduced to make the covariance $P_i$ increase at the initial stage.
In particular, $\lambda(l)=\frac{b}{l+1}$ with $b>0$ is chosen, such that the covariance of $P_i$ increases at the initial stage, and the rate of increase becomes slower as $l$ increases.
The third term on the right-hand-side of (\ref{eq:local:estimator}) is introduced to pursue the consensus of the parameter estimates given by different local filters by taking into account the difference between the parameter estimate given by each filter and an average of the estimates provided by the other filters that are accounted for by the third term on the right-hand-side of (\ref{eq:local:estimator}).
Therefore, the same level of significance is assigned to the sensor nodes and the estimates given by the filters of the DEKF scheme.
For this consensus term, we determine that $\mu(l) = \frac{a}{{(l+1)}^{\chi_a}}$ where $a>0$ and $0<\chi_a<1$. In this way, the design can enable the estimates provided by different filters to reach a consensus at the initial period. Also, after a certain time period when a consensus has (almost) been reached, the contribution of the corresponding consensus term in (\ref{eq:local:estimator}) to the estimate of each filter decreases as $l$ increases.

\subsection{Implementation algorithm of the consensus-based DEKF}
The local filters are implemented recursively until the parameter estimates given by different filters have reached a consensus and have converged.
An algorithm is presented in this section to show the implementation steps for the proposed consensus-based DEKF.


\begin{alg} \label{alg:1}
Parameter estimation using the consensus-based DEKF.\vspace{1.5mm}

For the $i$th node, $i=1,\ldots,n$, do the following:
\begin{enumerate} \itemsep 0em

    \item  Set $k=0$.

    \item Both the soil moisture and the capillary pressure sensors collect measurements and send them to the corresponding filter of the DEKF. \vspace{1mm}
%

    \item  Perform random permutation of the available $k+1$ pairs of measurements.

    \item The $i$th filter is initialized with an initial guess of the unknown parameters. \vspace{1mm}


    \item {\bf for} $l=0,1,\ldots,k$, do:
    \begin{enumerate}
        \item [5.1] The $i$th filter requests and receives $\hat{ \beta}_j^{(k)}(l)$ for $j\in\mathbb I\setminus\left\{i\right\}$;
        \item [5.2] The $i$th filter is evaluated following (\ref{eq:local:estimator}) to provide $\hat{\beta}_i^{(k)}(l+1)$;
    \end{enumerate}

    \item Set $ \hat{\beta}^i(k) = \hat{\beta}_i^{(k)}(k+1)$.

    \item Set $k = k+1$ and go to Step 2.\vspace{1.5mm}

\end{enumerate}
\end{alg}
In the above algorithm, $\hat{\beta}^i(k)$ denotes the final estimate of $\beta$ given by the $i$th filter, $i\in\mathbb I$, at sampling time $k$, and can be used as an estimate of $\beta$ if it is determined that the estimates provided by the filters have converged and a consensus has been reached.

The estimate provided by any filter $i$, $i\in\mathbb I$, can be used as the estimate of $\beta$ by consensus-based DMHE for state estimation after the estimates given by the filters have reached a consensus and have converged. This can be determined by calculating the Euclidean norms of the difference between the estimates given by two filters and the difference between the estimates given by each filter at two consecutive sampling times. If the norms are sufficiently small, the parameter estimates of the consensus-based DEKF can be used in state estimation.


\begin{rmk}
According to Algorithm~\ref{alg:1}, at each sampling time, the previous sensor measurements are permuted and used repeatedly for calculating the parameter estimates recursively. This is because that the sampling interval is in general comparatively large for the agro-hydrological systems, and this treatment (permuting the measurements and evaluating the DEKF repeatedly at each sampling time) can facilitate the convergence of the parameter estimates when the number of measurements available to the DEKF is not sufficiently large.
\end{rmk}

\begin{rmk}
This consensus-based DEKF is still computationally efficient so that the online implementation is not difficult to realize.
Note that the number of measurements available to DEKF increases and the computational complexity for each evaluation step grows linearly with time (the number of available measurement pairs).
For cases when the DEKF is implemented for a long time period and the number of measurements becomes too large, the earliest pair of measurements can be discarded when a new pair is available. In this way, the number of measurement pairs used by DEKF is limited, which ensures the computation is tractable.
\end{rmk}


\section{Consensus-based distributed MHE}

In this section, we present the consensus-based distributed MHE design that is primarily for state estimation. This distributed MHE takes advantage of the soil measurements from the sensors as well as the estimates of parameters in $\beta$ provided by DEKF to online estimate the full state information and the remaining parameter $K_{sat}$.

\subsection{The local MHE estimator design}

For each sensor node, we design a local consensus-based MHE estimator. Each local estimator provides an estimate of the full-state of system (\ref{eq:state:space}) and an estimate of the remaining unknown parameter $K_{sat}$. Similar to the DEKF scheme, the local estimators exchange information in terms of the estimates of the system state and the parameter with each other. These local estimators are required to be evaluated iteratively at each sampling time, and they will achieve a consensus on their estimates, which will be further used to obtain an estimate of the soil moisture information at different locations of the soil profile.

Specifically, for the $i$th sensor node, $i\in\mathbb I$, and at the $k$th sampling time,
the consensus-based MHE estimator is formulated as in the following form:

\begin{subequations}\label{pcc:eqn:mhe:0}
\begin{align}
& \qquad\qquad\qquad\qquad \min\limits_{{\tilde K}^{i,p}_{sat},\left\{ \tilde {x}^{i,p}(d|k)\right\}_{d=k-N}^k} J_i\left(k,N,{\tilde K}^{i,p}_{sat},{\bar K}^{i,p-1}_{sat},{\tilde {x}^{i,p}},{{\bar x}^{i,p-1}}, \tilde w^{i,p},\tilde v^p_{i}\right) \label{pcc:eqn:mhe:cost} \\[0.23em]
&~\quad\quad\textmd{\bf{ s.t.~}}\quad~ {\tilde x}^{i,p}(d+1|k)= f\left({\tilde x}^{i,p}(d|k), {\tilde K}^{i,p}_{sat}, {\hat\beta}^{i}, u(d)\right) + {\tilde w}^{i,p}(d),~~d=k-N,\ldots,k-1\label{pcc:eqn:mhe:model_x}\\
& ~\qquad\quad\quad\quad\quad\qquad~~ \tilde v^p_{i}(d) = y_{i}(d) - g_i\left({\tilde x}^{i,p}(d|k),{\hat\beta}^{i}\right),~~d=k-N,\ldots,k \label{pcc:eqn:mhe:model_y} \\[0.23em]
& ~\quad\quad\quad\quad\quad\quad\quad~~~~~~ \tilde{x}^{i,p}(d) \in {\m X},\;~~d=k-N,\ldots,k \label{pcc:eqn:mhe:cons1}\\[0.23em]
& ~\quad\quad\quad\quad\quad\quad\quad~~~~~ \tilde w^{i,p}(d) \in {\m W},\;~~d=k-N,\ldots,k-1 \label{pcc:eqn:mhe:cons2}\\[0.23em]
& ~\quad\quad\quad\quad\quad\quad\quad~~~~~~\tilde v^p_{i}(d) \in {\m V}_{i},\;~~~d=k-N,\ldots,k \label{pcc:eqn:mhe:cons3}
\end{align}
\end{subequations}
In (\ref{pcc:eqn:mhe:0}), $J_i$ is the individual cost function for the $i$th estimator and will be described in detail later, $N$ denotes the length of the moving estimation window,
$\tilde x^{i,p}(d|k)$ represents the estimate of the full state $x(d)$ obtained by the $i$th estimator in the $p$th iteration, $d=k-N,\ldots,k$, determined in the $p$-th iteration step at time $k$, $\tilde w^{i,p}$ is the estimate of the process disturbance given by the $i$th local estimator determined in the $p$th iteration step, and $\tilde v_i^p$ is an estimate of the measurement noise of the $i$th sensor node in the $p$th iteration step.

In the optimization problem (\ref{pcc:eqn:mhe:0}), (\ref{pcc:eqn:mhe:cost}) is the objective that seeks the optimal state estimation sequence $\left\{\tilde {x}^{i,p}(d|k)\right\}_{d=k-N}^{k}$ in the $p$th iteration step by minimizing the consensus-based individual cost $J_i$, (\ref{pcc:eqn:mhe:model_x}) and (\ref{pcc:eqn:mhe:model_y}) serve as model constraints,  and (\ref{pcc:eqn:mhe:cons1}) to (\ref{pcc:eqn:mhe:cons3}) impose hard constraints on the estimates of system state, the process disturbance and measurement noise, respectively.
From time $k$ to ${k+1}$, the MHE-based estimator for sensor node $i$ in (\ref{pcc:eqn:mhe:0}) is evaluated iteratively for at least once to obtain its optimal state estimate sequence for the entire process, i.e., $\left\{\tilde {x}^{i,p}(d|k)\right\}_{d=k-N}^{k}$. From time $k$ to $k+1$, the state estimate sequence obtained in the last iteration step is treated as the optimal estimate sequence at this sampling time, and is denoted by $\left\{{\hat x}^i(d|k)\right\}_{d=k-N}^k$. Its last element serves as the optimal estimate of actual state $x(k)$ provided by the estimator for the $i$th sensor node.

In the objective function (\ref{pcc:eqn:mhe:cost}), the cost function $J_i$ is with the following form:
\begin{equation}
J_i\left(k,N,{\tilde K}^{i,p}_{sat}, {\bar K}^{i,p-1}_{sat},{\tilde {x}^{i,p}},{{\bar x}^{i,p-1}}, \tilde w^{i,p},\tilde v^p_{i}\right) = \sum\limits_{d=k-N}^{k-1} \big|\tilde w^{i,p}(d|k)\big|^2_{Q^{-1}} +   \sum\limits_{d=k-N}^{k} \big|\tilde v^p_{i}(d|k)\big|^2_{R_{i}^{-1}} + V^{i,p}(k-N)
\end{equation}
where
$Q$ and $R_i$ are selected to be positive definite matrices, and $V_i(t_{k-N})$ represents the initial cost that consists of two parts as follows:
\begin{equation}\label{eq:cost:arrive}
\begin{aligned}
V^{i,p}\left(k-N\right)& = L^{i,p}(k-N)  + C^{i,p}(k-N)
\end{aligned}
\end{equation}
In (\ref{eq:cost:arrive}), the first term on the right-hand-side is the initial cost that penalizes the difference between the {\em a posteriori} estimates of the state and the parameter obtained at different sampling times, and is defined as follows:
\begin{equation}\label{eq:cost:arrive:individual}
\begin{aligned}
L^{i,p}(k-N) &= \Big\|\tilde {x}^{i,p}(k-N|k)-\hat { x}^{i}{(k-N|k-1)}\Big\|_{{\Pi_{L,i}}^{-1}}+\Big\|{\tilde K}^{i,p}_{sat}(k)-{\hat K}^{i}_{sat}(k-1)\Big\|_{\mu_{L,i}}
\end{aligned}
\end{equation}
where $\hat { x}^{i}{(k-N|k-1)}$ is the optimal estimate of the state at $k-N$ made by the $i$th estimator at sampling time $k-1$, $\Pi_{L,i}$ is a weighting matrix that should be made positive definite and bounded, and $\mu_{L,i}$ is a tuning parameter that needs to be made positive.

The second term on the right-hand-side of (\ref{eq:cost:arrive}) accounts for the consensus-based cost as adapted from \cite{farina2012distributed}. It penalizes the difference between the estimates given by the local estimator and a weighted sum of the estimates given by the other estimators in the belief that the use of consensus can facilitate the convergence of the estimates of the states and the parameter to the actual values and improve the robustness of the obtained estimates. Specifically, the consensus-based cost is defined as:
\begin{equation}\label{eq:cost:arrive:consensus}
\begin{aligned}
C^{i,p}(k-N) &= \Big\|\tilde {x}^{i,p}(k-N|k)-{\bar x}^{i,p-1}_{k-N}\Big\|_{{\Pi_{C,i}}^{-1}}+ \Big\|{\tilde K}^{i,p}_{sat}(k)-{\bar K}^{i,p-1}_{sat}\Big\|_{\mu_{C,i}}
\end{aligned}
\end{equation}
where $\Pi_{C,i}$ is a positive definite matrix and $\mu_{C,i}$ is a scalar that needs to be made positive, $ {\bar x}^{i,p-1}_{k-N}$ represents an average of the state estimates and ${\bar K}^{i,p-1}_{sat}$ represents an average of the parameter estimates given by the estimators other than the $i$th estimator:
\begin{equation}\label{eq:cost:arrive:barx}
{\bar x}^{i,p-1}_{k-N} = \left\{ \begin{array}{l}
\displaystyle\sum_{j\in\mathbb I\setminus\left\{i\right\}} k_{ij}\tilde { x}^{j,p-1}(k-N|k),~~~\text{if~$p>1$}\\[0.36em]
\displaystyle\sum_{j\in\mathbb I\setminus\left\{i\right\}} k_{ij}\hat { x}^{j}(k-N|k-1),~~\text{if~$p=1$}
\end{array} \right.
\end{equation}
and
\begin{equation}\label{eq:cost:arrive:ksat}
{\bar K}^{i,p-1}_{sat} = \left\{ \begin{array}{l}
\displaystyle\sum_{j\in\mathbb I\setminus\left\{i\right\}} k_{ij}{\tilde  K}^{j,p-1}_{sat}(k),~~~~~\text{if~$p>1$}\\[0.36em]
\displaystyle\sum_{j\in\mathbb I\setminus\left\{i\right\}} k_{ij}{\hat K}^{j}_{sat}(k-1),~~\text{if~$p=1$}
\end{array} \right.
\end{equation}

An empirical solution to the selection of the weighting matrices $\Pi_{L,i}$ $\&$ $\Pi_{C,i}$ and the tuning parameters $\mu_{L,i}$ $\&$ $\mu_{C,i}$ in (\ref{eq:cost:arrive:individual}) and (\ref{eq:cost:arrive:consensus}) is to assign constant values to them, which has been considered as a common treatment in the nonlinear context; see, for example, \cite{farina2012distributed}. The selection of these weighting matrices and scalars for the irrigation process is conducted and discussed in Section~\ref{section:setting}. In (\ref{eq:cost:arrive:barx}) and (\ref{eq:cost:arrive:ksat}), $k_{ij}$ are weighting parameters that satisfy $\sum_{j\in\mathbb I\setminus\left\{i\right\}}k_{ij}=1$. A practical choice is to set $k_{ij}=\frac{1}{n_y-1}$, which implies that the same weight is assigned to all the estimators of DMHE in the consensus-based cost.

\begin{rmk}
Within the first $N$ sampling periods since DMHE is activated, the estimation window is not sufficiently long, and the estimators are evaluated in a full-information mode. The full-information design for local estimator $i$ of the DMHE is also based on (\ref{pcc:eqn:mhe:0}), while the initial cost in (\ref{eq:cost:arrive:individual}) is excluded from the objective function in this case.
\end{rmk}

\subsection{Iterative distributed estimation algorithm}
In this subsection, an algorithm is presented to illustrate how each local estimator of the consensus-based DMHE scheme is implemented.

\begin{alg} \label{soil:alg:2}
State estimation using the consensus-based DMHE

\hspace{-7mm}At each sampling time $k \geq\tau_0+N$, set $p=1$, and carry out the following steps:\vspace{-1mm}

\begin{enumerate} \itemsep 0em

        \item[1.]
          Estimator $i$, $i\in\mathbb I$, receives the measurement $y_i(k)$ and the updated estimate $\hat{\beta}^i(k)$ from the $i$th filter of DEKF.
        \item[2.] {\emph{\textbf{if}}} $p=1$, {\emph{\textbf{then}}}

        ~~~Estimator $i$ receives ${\hat x}^{j}(k-N|k-1)$ and $\hat K^{j}_{sat}(k-1)$ from each  estimator $j$, $j\in\mathbb I\setminus\left\{i\right\}$.

        {\emph{\textbf{else}}}, do:

        ~~~Estimator $i$ receives ${\tilde x}^{j,p-1}(k-N|k)$ and $\tilde K^{j,p-1}_{sat}(k)$ from each estimator $j$, $j\in\mathbb I\setminus\left\{i\right\}$.

        \item[3.]
          Based on $\left\{y_i\right\}^{k}_{k-N}$ and the state and parameter estimates given by the other estimators, estimator $i$ is evaluated following (\ref{pcc:eqn:mhe:0}) to provide state estimate $\left\{\tilde{x}^{i,p}(d|k)\right\}_{d=k-N}^k$ and parameter estimate ${\tilde K}^{i,p}_{sat}(k)$, $i\in \mathbb I$.

        \item[4.] {\emph{\textbf{if}}} $p < p_{\max}$,  {\emph{\textbf{then}}}
             \newline
           \text{~~~}set $p=p+1$, and go to step 2.
             \newline
             {\emph{\textbf{else}}}, do:
             \newline
           \text{~~~}go to step 5.

        \item[5.] Use $\hat x^i(k)=\tilde x^{i,p}(k|k)$ and ${\hat K}^{i}_{sat}(k) = {\tilde  K}^{i,p}_{sat}(k)$ as the estimates of the full-state and the parameter given by estimator $i$, $i\in \mathbb I$. Set $k=k+1$ and go to step 1.
%

\end{enumerate}
\end{alg}

In Algorithm~\ref{soil:alg:2}, $\tau_0$ is the sampling time when the estimates of $\beta$ given by the DEKF have converged. From this sampling time, the estimates of $\beta$ are treated as reliable parameter estimates, the DMHE is activated at $\tau_0$, and each of the estimators gives a full-state estimate and an estimate of the remaining model parameter $K_{sat}$. $p_{\max}$ is the maximum number of iteration steps at each sampling time, which indicates that the estimators of the DMHE are evaluated iteratively for $p_{\max}$ times during each sampling time.
Algorithm~\ref{soil:alg:2} describes the implementation steps after the estimators are evaluated in a moving horizon manner. At each sampling time $k$, $\tau_0\leq k<\tau_0+N$, the estimators are evaluated based on full information with all the estimators being initialized at $\tau_0$.

\begin{rmk}
In this design, the number of iterative evaluation steps for the consensus-based DMHE is fixed. Alternatively, a triggering condition can be imposed to govern the iterative evaluation. In particular, when the estimates given by each estimator in two consecutive iteration steps become close to each other, then the iterative evaluation can be terminated. Based on this treatment, computational complexity may be reduced as the number of iteration steps may become small after the estimates have converged. More discussions with respect to this point can be found in \cite{YDL2018CHERD}.
\end{rmk}



\section{Recovery of soil moisture and key steps for the entire mechanism}\label{section:recovery}

Our objective is to obtain good and reliable estimates of the soil moisture information at all the discretized nodes. This objective can be achieved based on the estimates of the model parameters and the system states in terms of soil water pressure head provided by the two consensus-based schemes. Specifically, let us recall the soil-water retention equation in (\ref{eq:retention}). This equation can be applied to each of the discretized nodes in the soil, such that in the $i$th compartment, the relationship between the soil moisture and the soil water pressure head as well as the four parameters involved in $\beta$ is characterized by:
\begin{equation}\label{eq:retention:discrete:node:i}
\theta_i =
     \left(\theta_{s}-\theta_{r}\right)\left(1+(-\alpha h_i)^n\right)^{-\left(1-\frac{1}{n}\right)}+\theta_{r},~~~i=1\ldots,n_x
\end{equation}
By substituting $h_i$ and the four parameters in (\ref{eq:retention:discrete:node:i}) with their estimates given by the consensus-based mechanism, an estimate of the soil moisture in the $i$th compartment as denoted by $\hat\theta_i$ can be obtained.

Now, we are in a position to summarize the key steps of the entire consensus-based mechanism described by the flowchart in Figure~\ref{flowchart_propose_design}:

\begin{enumerate}
    \item The DEKF is evaluated following Algorithm~\ref{alg:1} to give $\hat{ \beta}^i(k)$, $i\in \mathbb I$, at every sampling time.
    \item When the parameter estimates given by the filters of the consensus-based DEKF have reached a consensus and have converged, the time instant is labeled as $\tau_0$. From $\tau_0$, the DMHE becomes active.
    \item For $k\geq \tau_0$, do the following:
     \begin{itemize}
        \item the DEKF sends $\hat{\beta}^i(k)$, $i\in \mathbb I$, to the estimators of the DMHE;
        \item the DMHE is executed following Algorithm~\ref{soil:alg:2} to provide $\hat x^i(k)$ and ${\hat K}^{i}_{sat}(k)$;
        \item (\ref{eq:retention:discrete:node:i}) is used to recover the soil moisture at different locations of the soil profile based on $\hat x^i(k)$ and $\hat{\beta}^i(k)$, $i\in\mathbb I$.
     \end{itemize}
\end{enumerate}

Considering the fact that the unknown parameters may change over time, the tensiometers will continue to collect pressure measurements after $\tau_0$, so that the DEKF scheme can provide undated estimates of the unknown parameters when necessary. Once the updated estimates of the parameters are available, they will be sent to and are used by the DMHE.

\begin{rmk}
Note that in certain cases, the parameters of the agro-hydrological system may only change slowly. In these cases, it is not necessary to require that the tensiometers collect head pressure measurements and the DEKF is evaluated every sampling time. Instead, one more step can be added to Step 3 of the above procedure, which is to deactivate the DEKF and the tensiometers once a new/updated estimate of parameter vector $\beta$ is provided by the DEKF, and re-activates them after a prescribed time period or when the model residuals for the retention equation in (\ref{eq:state:space:local:2}) become large. In this way, the consumption of both computing resource and energy can be reduced.
\end{rmk}

\section{Simulation results}\label{section:setting}

\subsection{Irrigation system settings}
\begin{table}[t]
\caption{The actual values of the five parameters of the soil profile}\vspace{2mm}
\label{agri:parameter:tab0}\renewcommand\arraystretch{1.39}
\newcommand{\tabincell}[2]{\begin{tabular}{@{}#1@{}}#2\end{tabular}}
  \centering
  \begin{tabular}{c|c|c|c|c|c}\hline\hline
		& $K_{sat}$ &$\theta_{s}$ &$\theta_{r}$ & $\alpha$ & $n$\\
\hline
		Loam & $2.89\times 10^{-6}~\text{m}/\text{s}$  & 0.430 $\text{m}/\text{m}$ & 0.0780~ $\text{m}/\text{m}$ & 3.60$/\text{m}$ &1.56\\
\hline\hline
\end{tabular}
\end{table}

Without loss of generality, we consider a loam soil profile with a depth of 67 cm. In terms of spatial discretization, we determine to have 32 states (i.e., $n_x=32$); this ensures that we can achieve an accurate numerical approximation of the Richards equation in (\ref{eq:richards}). Consequently, the entire profile is partitioned into 32 small compartments vertically, while the soil moisture in each compartment is considered to be homogeneous.

\begin{figure}[t]
\centerline{\includegraphics[width=0.25\textwidth]{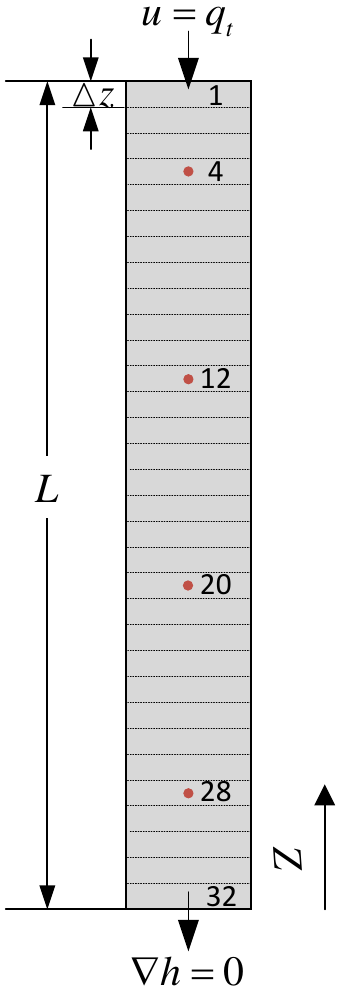}}
\caption{The 32 compartments of the soil profile after space discretization and selected sensor nodes (the compartments with red dots are treated as sensor nodes).}
\label{figure:soil:profile}
\end{figure}
The soil water pressure head in the small compartments constitutes the state vector $x$ of system (\ref{eq:state:space:local}). On the surface of the top compartment, water flow that enters the soil profile is at a rate of $1.944\times 10^{-3}~ \text{m}/\text{hr}$ for eight hours on a daily basis. At the bottom of the soil profile, we consider the case of zero capillary pressure head gradient, i.e., the free drainage boundary condition is applied to the bottom boundary. The actual values of the five parameters, which are the same as the parameters reported in \cite{carsel1988developing}, are presented in Table~\ref{agri:parameter:tab0}.

Among the 32 compartments, four compartments are selected as sensor nodes (i.e., $n_y=4$), and a tensiometer and a soil moisture sensor are deployed at each of the four sensor nodes. As shown in Figure~\ref{figure:soil:profile}, the four sensor nodes are equally spaced below the surface at 7.33 cm, 24.08 cm, 40.83 cm and 57.58 cm, respectively. This indicates that the sensors are placed at the center points of the 4th, 12th, 20th and 28th compartments, and these sensors can provide the measurements of the capillary pressure head and the soil moisture in these compartments.

\subsection {Simulation setting}
We consider synchronous sampling of all the sensor measurements. It is assumed that the sensors are able to provide sampled measurement every $4~\text{min}$. It is also assumed that the measurements are immediately available to the associated local filter/local estimator after they are collected by sensors, which implies that communication delay is not considered. This is a reasonable assumption considering the comparatively large sampling period.

There are four sensor nodes, such that there are four local filters in the consensus-based DEKF scheme and four local estimators in the consensus-based DMHE scheme. In terms of the DEKF, the initial guesses of the four parameters in $\beta$ are determined and shown in Table~\ref{agri:parameter:tab0:initial:guess}. It is worth mentioning that the initial guesses selected for the filters of the DEKF are significantly different from the actual values, since we aim to demonstrate the ability of the proposed consensus-based algorithm to provide parameter estimates subject to poor initial guesses. For (\ref{eq:local:estimator:2:a}), $P_i$ is made as $P_i(0) = \text{diag}\left(\left[1, 0.16, 15, 3\right]\right)\times \text{rand}(1)$ for $i=1,2,3,4$ where $\text{rand}(1)$ returns a single uniformly distributed random number between 0 and 1, and $Q$ and $R$ are made constant as $Q = \text{diag}\left(\left[0.0225, 0.0225, 0.0225, 0.0225\right]\right)$ and $R = 0.01$. Note that the primary requirement on them is that they should be made positive definite, and it is found that different choices of positive definite $Q$ and $R$ do not affect the estimation accuracy for the model parameters significantly.
In the filter (\ref{eq:local:estimator}), we adopt $\mu(l) = \frac{0.3}{{(l+1)}^{0.1}}$, such that the contribution of the corresponding innovation term in (\ref{eq:local:estimator}) is reduced as $l$ increases. In addition, we determine that $\lambda(l)=\frac{0.5}{l+1}$, thus $P_i$ increases more slowly as $l$ increases. The maximum number of measurements that can be used by the DEKF is set to be $k_{\max}=1000$.


\begin{table}[t]
\caption{The initial guesses for the unknown parameters selected for the local filters of the DEKF}\vspace{2mm}
\label{agri:parameter:tab0:initial:guess}\renewcommand\arraystretch{1.39}
\newcommand{\tabincell}[2]{\begin{tabular}{@{}#1@{}}#2\end{tabular}}
  \centering
  \begin{tabular}{c|c|c|c|c}\hline\hline
		& ${\hat\theta}_{s}$ &${\hat\theta}_{r}$ & $\hat\alpha$ & $\hat n$\\
\hline
		{\tabincell{c}{Initial guess \\[-0.1em]for Filter 1}}  & 0.33 $\text{m}/\text{m}$ & 0.37~ $\text{m}/\text{m}$ & 1.60$/\text{m}$ &0.57\\
\hline
		{\tabincell{c}{Initial guess \\[-0.1em]for Filter 2}}  & 0.48 $\text{m}/\text{m}$ & 0.12~ $\text{m}/\text{m}$ & 3.20$/\text{m}$ &1.36\\
\hline
		{\tabincell{c}{Initial guess \\[-0.1em]for Filter 3}}  & 0.73 $\text{m}/\text{m}$ & 0.15~ $\text{m}/\text{m}$ & 3.85$/\text{m}$ &1.87\\
\hline
		{\tabincell{c}{Initial guess \\[-0.1em]for Filter 4}}  & 0.62 $\text{m}/\text{m}$ & 0.04~ $\text{m}/\text{m}$ & 2.65$/\text{m}$ &0.70\\
\hline\hline
\end{tabular}
\end{table}

\begin{figure}[t]
\centerline{\includegraphics[width=0.6\textwidth]{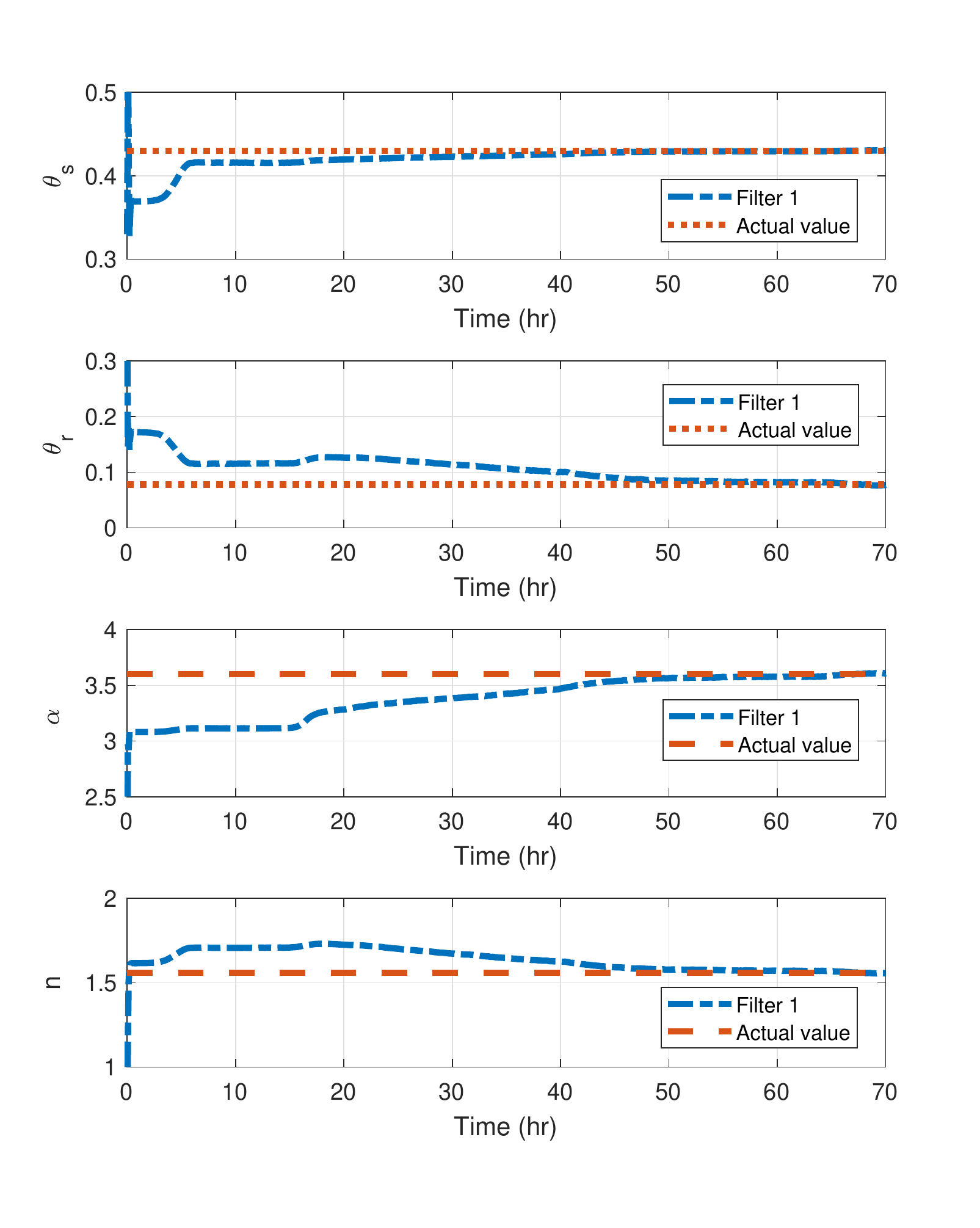}}
\caption{The parameter estimates provided by Filter 1 of the consensus-based DEKF and the actual values of the parameters.}
\label{figure:para:estimate:DEKF}
\end{figure}

For the DMHE scheme, the length of the estimation window is set to be $N=12$, and the weighting matrices/scalars are set as follows: $Q_i = {I}_{32}$, $R_i = 1$, $\Pi_{L,i} = \Pi_{C,i}= 0.6\times {I}_{32}$, $\mu_{L,i} = \mu_{C,i} = 1.0\times 10^{11}$, $i=1,2,3,4$.
It is required that each estimator of the DMHE is evaluated twice during each sampling time.
The sampling time for the DMHE is set to be $20~\text{min}$.
Additive process disturbance and random measure noise are generated following normal distribution with zero mean.

The initial states at all the nodes of the soil profile are set to be $-0.5139~\text{m}$. The initial guesses for the states used by the four local estimators of the DMHE are picked as $0.6 x(0)$, $1.5 x(0)$, $0.5 x(0)$ and $2.3 x(0)$, respectively, where $x(0)$ denotes the initial condition of the process. The initial guesses of the remaining parameter $K_{sat}$ used by the estimators of the DMHE are selected in the same way.
In the local estimators of the DMHE, hard constraints which are shown in Table~\ref{agri:parameter:tab1:constraint} are imposed on the process states and the remaining parameter $K_{sat}$ that also needs to be estimated. These constraints account for, for example, practical case scenarios that the states that represent capillary pressure head take values within $[-1, 0)$ and the parameter $K_{sat}$ is positive.
Note that as compared to \cite{Bo2020}, the initial guesses of the states and the parameter are made much further away from the actual values.
Also, the lower and upper bounds on $K_{sat}$ together account for a much looser constraint on this parameter in DMHE as compared to the centralized MHE design in \cite{Bo2020}, aiming to demonstrate the efficacy of the proposed method in the cases when much less prior information is available.

\begin{figure}[t]
\centerline{\includegraphics[width=0.8\textwidth]{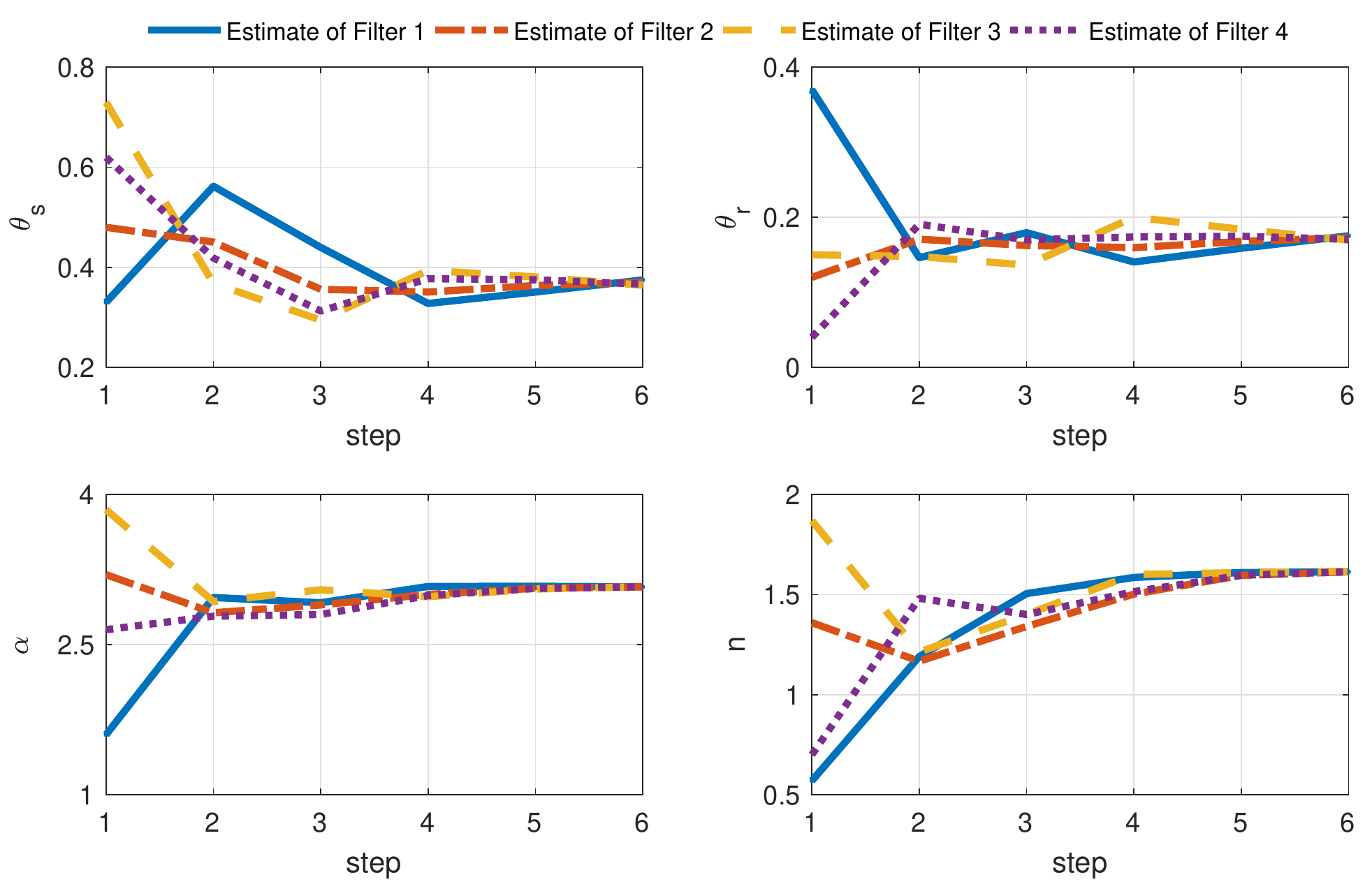}}
\caption{The parameter estimates given by the local filters of the DEKF at the initial stage.}
\label{figure:para:estimate:DEKF:initial:steps}
\end{figure}

\begin{table}
	\caption{Lower and upper bounds used in consensus-based DMHE}\vspace{2mm}
	\centering \label{agri:parameter:tab1:constraint}\renewcommand\arraystretch{1.39}
	\begin{tabular}{c | c | c }
\hline\hline
		& $\hat{x}$ $(\text{m})$	& $\hat{K}_{sat}$ $(\text{m}/\text{s})$ \\
\hline
		Lower bounds	    & -1.00  & $1.0\times 10^{-7}$ \\
\hline
		Upper bounds	&  $-1.00\times 10^{-6}$  &$1.0\times 10^{-5}$  \\
\hline\hline		
	\end{tabular}
\end{table}

\begin{figure}[t]
\centerline{\includegraphics[width=0.58\textwidth]{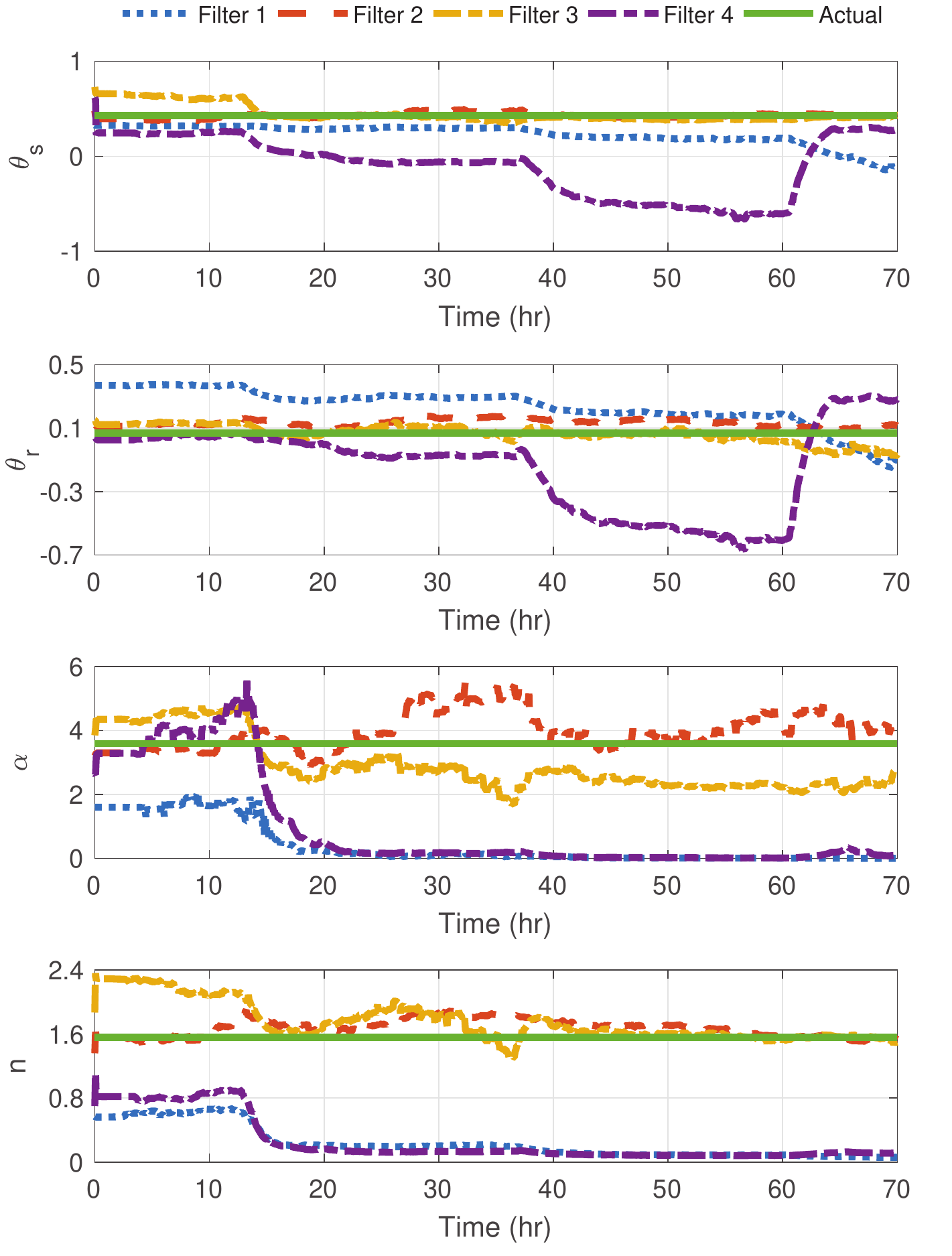}}
\caption{The estimates of the parameters provided by extended Kalman filtering without consensus (Cases 1 to 4 are based on the same Kalman filtering algorithm yet different initial guesses as listed in Table~\ref{agri:parameter:tab0:initial:guess}).}
\label{figure:para:estimate:CEKF}
\end{figure}
\subsection{Results}\label{section:results}
\subsubsection{Parameter estimates of the DEKF}

First, the DEKF scheme is utilized to estimate the four parameters in $ \beta$. The trajectories of the estimates of the parameters $\theta_{s}$, $\theta_{r}$, $\alpha$ and $n$ provided by the four local filters are shown in Figure~\ref{figure:para:estimate:DEKF}.
The estimates of the parameters given by all the four local filters can eventually converge and stay close to the actual parameters despite the inaccurate initial guesses. In addition, the trajectories of the parameter estimates of the local filters at the initial stage are given in Figure~\ref{figure:para:estimate:DEKF:initial:steps}. It indicates that the parameter estimates of the four filters become very close to each other very quickly owing to the consensus-based design and the selected parameter $\mu(l)$ in  (\ref{eq:local:estimator}). Based on the estimates of the filters, it is determined that the estimates have converged approximately 50 hours after the initial time instant by examining the differences between the estimates of the filters. Therefore, the parameter estimates of the DEKF can be used by the DMHE, and the DMHE can be activated to give estimates of the state and the remaining parameter 50 hours after the DEKF was activated.

The proposed consensus-based DEKF is also compared to centralized extended Kalman filtering (EKF) to demonstrate the superiority when the initial guess is distant from the actual value.
In particular, we consider four different cases when centralized EKF is initialized with different initial guesses as given in Table~\ref{agri:parameter:tab0:initial:guess}. These initial guesses are also used by the local filters of the DEKF scheme. The parameter estimation results are presented in Figure~\ref{figure:para:estimate:CEKF}. From the results, for each of the four cases, the estimates for at least one of the parameters cannot converge to the corresponding actual values as the poor initial guesses in Table~\ref{agri:parameter:tab0:initial:guess} are used.
The above comparison implies that, in the absence of a good initial guess and in the presence of noise, the proposed consensus-based DEKF can be more desirable due to its capability of providing convergent and more reliable parameter estimates.

\subsubsection{Results of the DMHE and the soil moisture estimates}

\begin{figure}[t]
\centerline{\includegraphics[width=0.72\textwidth]{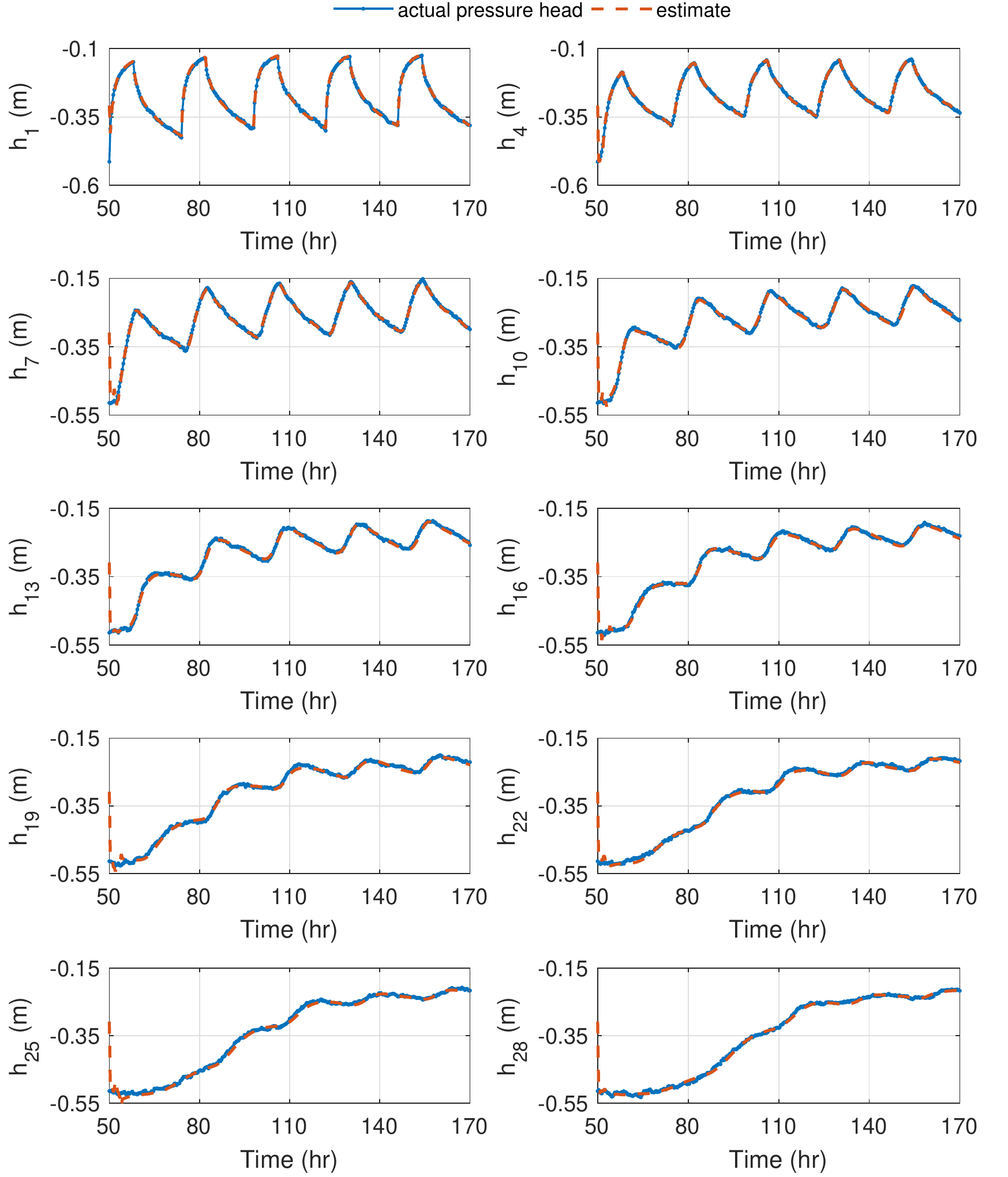}}
\caption{The trajectories of the actual water pressure head and the estimates provided by the DMHE at different depths of the soil profile.}
\label{figure:para:estimate:DMHE}
\end{figure}

The parameter estimates provided by the DEKF are sent to the DMHE for the estimation of the states and the remaining model parameter $K_{sat}$.
Simulations are carried out based on the DMHE, and the estimates of the states representing the water pressure head in different compartments of the soil profile are presented in Figure~\ref{figure:para:estimate:DMHE}. The DMHE also consists of four local estimators. In this figure, only the estimates provided by the first local estimator are given, while the estimates provided by the other local estimators are omitted since their estimate trajectories tend to overlap with each other shortly at the initial stage. The results confirm that the proposed DMHE can provide accurate estimates of the water pressure head based on the parameter estimates provided by the DEKF, even when the initial guesses for the local estimators are significantly different from the initial state. After having converged, the average of the estimate ${\hat K}_{sat}$ given by the DMHE is $2.77\times 10^{-6}~\text{m}/\text{s}$, which is close to the actual value of $K_{sat}$. It is worth noting that the accuracy of soil moisture estimates given by the proposed mechanism is independent of the quality of the estimate of $K_{sat}$ according to Section~\ref{section:recovery}. Specifically, the calculation of the soil moisture estimates only requires the estimates of the water pressure head and the four parameters in $\beta$, while $ K_{sat}$ is not needed.

\begin{figure}[!t]
\centerline{\includegraphics[width=0.72\textwidth]{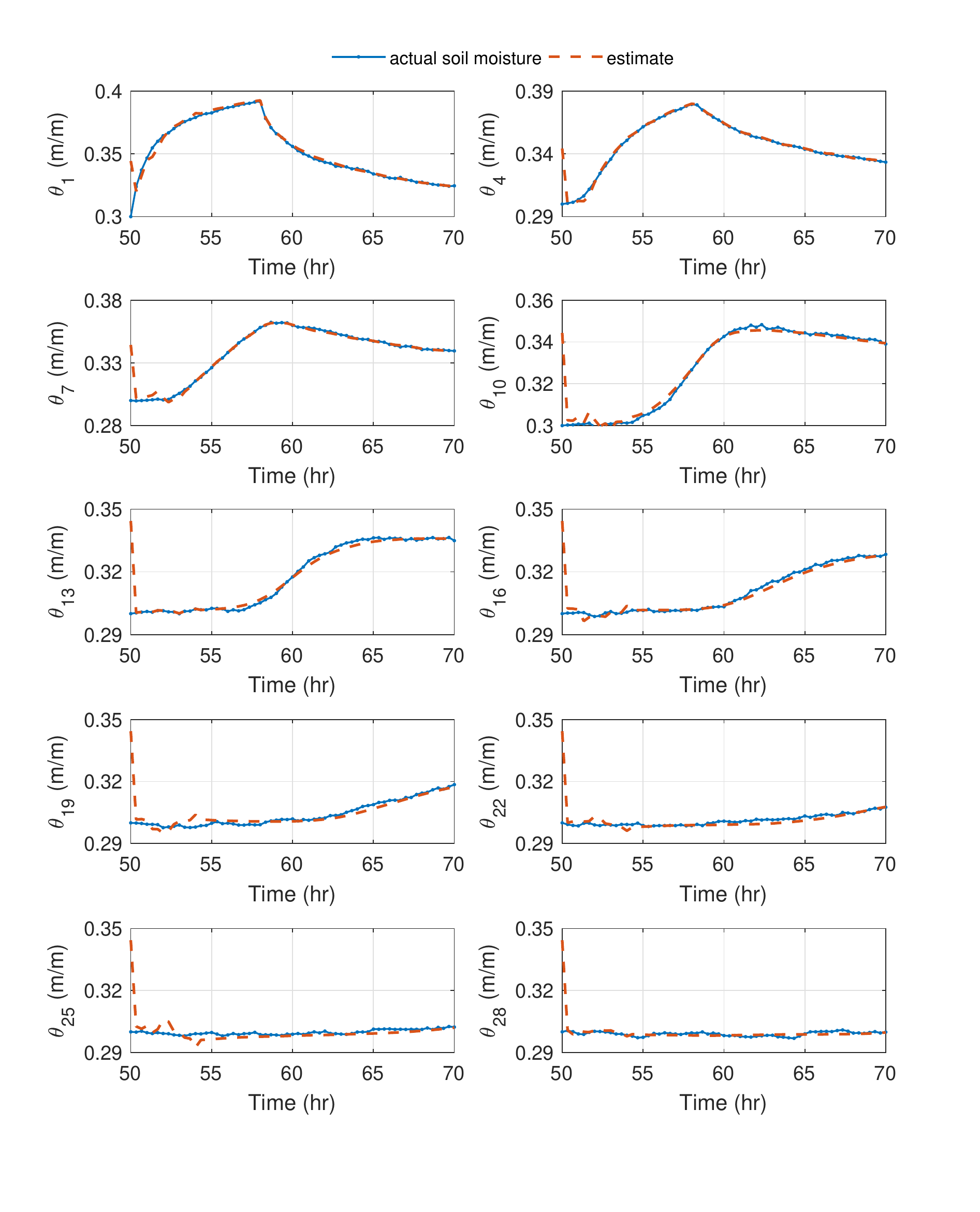}}
\caption{The trajectories of the actual soil moisture and the estimates of the soil moisture calculated based on the proposed consensus-based mechanism.}
\label{figure:para:estimate:soil:moisture}
\end{figure}

Next, the parameter estimates provided by the DEKF and the estimates of the soil water pressure head are jointly used to re-construct the soil moisture at different depths of the soil profile based on (\ref{eq:retention:discrete:node:i}). Since the estimates given by the local filters/estimators have reached a consensus, the soil moisture estimates are calculated based on the data provided by Filter 1 of the DEKF and Estimator 1 of the DMHE. The estimates of the soil moisture are presented in Figure~\ref{figure:para:estimate:soil:moisture}. The trajectories of the soil moisture estimates at the selected depths track the trend of the actual moisture accurately, which confirms that the proposed consensus-based mechanism scheme can provide good estimates of the soil moisture information at different depths. Also, the results show that the estimates of the soil moisture converge quickly to the actual values, which is primarily owing to the use of consensus in local estimators. Note that the soil moisture can be recovered based on the estimates provided by the filter and the estimator associated with any sensor node.

Finally, we also compare the proposed distributed mechanism with a centralized MHE method proposed in \cite{Bo2020} that also treats an estimation problem for (\ref{eq:richards}). In the centralized MHE, the soil moisture measurements collected from the four sensor nodes are used. We consider four cases when the initial guesses for the centralized MHE are made the same as those used by Filter $i$ of the DEKF and Estimator $i$ of the DMHE, $i=1,2,3,4$. The size of the estimation window is also 12 for the centralized MHE, and process and measurement noise variances of the same magnitudes are considered. The remaining settings are determined in a way such that a fair comparison is made between the two methods. In these four cases, we aim to use the centralized MHE to estimate the parameters and the states simultaneously. However, the centralized method in \cite{Bo2020} fails to provide good estimates in any of the four cases. This fact implies that when initial guesses of the parameters and the states are inaccurate, it is necessary to apply the proposed mechanism as needed for providing reliable soil moisture estimates.

\begin{rmk}
In the simulations, the actual model parameters are set to be constant. It is worth noting that the proposed consensus-based DEKF is also capable of handling the cases when the unknown model parameters are time-varying.
We consider different sampling periods for the two distributed estimation schemes. It is required that the DEKF is evaluated more frequently, such that it can provide accurate parameter estimates to the DMHE in a timely manner, which is important for the DMHE to be activated at the initial stage.
\end{rmk}


\section{Conclusions}

In this paper, the soil moisture estimation problem was addressed for agro-hydrological systems. A consensus-based estimation mechanism that involves two schemes was proposed. Within the consensus-based framework,
a consensus-based EKF scheme was proposed to estimate four parameters related to the soil properties, and a consensus-based distributed MHE scheme was proposed to estimate the full-state of the system. The local filters/esitmators of the two schemes exchange information and are evaluated in a collaborative manner such that the parameters and the states are simultaneously estimated online. Soil moisture information in different compartments of the soil profile can be recovered based on the estimates provided by the consensus-based mechanism. The results confirmed that accurate soil moisture estimates can be obtained, and the proposed mechanism is much less sensitive to the quality of initial guesses, thus is more robust as compared to centralized methods.

\section*{Acknowledgment}

Financial support from Natural Sciences and Engineering Research Council of Canada is gratefully acknowledged.


\end{document}